\documentclass[
 superscriptaddress,
 reprint,
 amsmath, amssymb,
 aps,prd, preprintnumbers, nofootinbib, floatfix, longbibliography
]{revtex4-1}

\usepackage{graphicx}
\usepackage{dcolumn}
\usepackage{hyperref}
\usepackage{bm}
\usepackage{ulem}
\usepackage{color}
\usepackage{amssymb}
\usepackage{amsmath}
\usepackage{comment}
\usepackage{adjustbox}
\usepackage{rotating}
\usepackage{multirow}
\usepackage{graphicx}
\usepackage{lipsum}
\usepackage{slashed}
\begin{document}
\preprint{}
\title{Boosting indirect detection of a secluded dark matter sector }

\author{Jinmian Li}
\email{jmli@scu.edu.cn}
\affiliation{College of Physics, Sichuan University, Chengdu 610065, China}

\author{Takaaki Nomura}
\email{nomura@scu.edu.cn}
\affiliation{College of Physics, Sichuan University, Chengdu 610065, China}

\author{Junle Pei}
\email{peijunle@ihep.ac.cn}
\affiliation{Institute of High Energy Physics, Chinese Academy of Sciences, Beijing 100049, China}
\affiliation{Spallation Neutron Source Science Center, Dongguan 523803, China}

\author{Xiangwei Yin}
\email{yinxiangwei@itp.ac.cn}
\affiliation{CAS Key Laboratory of Theoretical Physics, Institute of Theoretical Physics, Chinese Academy of Sciences, Beijing 100190, China}
\affiliation{School of Physical Sciences, University of Chinese Academy of Sciences, No. 19A Yuquan Road, Beijing 100049, China}

\author{Cong Zhang}
\email{zhangcong.phy@gmail.com (corresponding author)}
\affiliation{College of Physics, Sichuan University, Chengdu 610065, China}
\affiliation{Bethe Center for Theoretical Physics and Physikalisches Institut, Universit\"{a}t Bonn, Nussallee 12, D-53115 Bonn, Germany}

%\date{\today}

%%%%%%%%%%%%%%%%%%%%%%%%%%%%%%%%%%%%%%%%%%%%%%%%%%%%%%%%%%%%%%%%%%%%%%%%%%%%%%%%%%%%%%%%

\begin{abstract}
Dark Matter (DM) residing in a secluded sector with suppressed portal interaction could evade direct detections and collider searches. The indirect detections provide the most robust probe to this scenario.  
Depending on the structure of the dark sector, novel DM annihilation spectra are possible. 
The dark shower is a common phenomenon {for particles in the dark sector which take part in strong interactions and are boosted.}
In terms of simplified two-component DM models with vector portal interaction and pseudoscalar portal interaction, we study the dark showering effects for DM indirect detection.  
In those models, the heavier DM component which dominates the relic density annihilates into boosted lighter species. 
Together with the large coupling through which the lighter DM annihilates away in the early universe, the showered spectra provide as the smoking gun for the DM existence. 
Considering bounds obtained by the AMS-02 positron data and Fermi-LAT measurement of gamma-ray from the dwarf galaxies, we find the dark shower could open a new region of sensitivity that can not be probed before.

\end{abstract}

\maketitle

%%%%%%%%%%%%%%%%%%%%%%%%%%%%%%%%%%%%%%%%%%%%%%%%%%%%%%%%%%%%%%%%%%%%%%%%%%%%%%%%%%%%%%%%

%%%%%%%%%%%%%%%%%%%%%%%%%%%%%%%%%%%%%%%%%%%%%%%%%%%%%%%%%%%%%%%%%%%%%%%%%%%%%%%%%%%%%%%%

%\noindent
%\paragraph{Introduction}
\section{Introduction}
%\label{sec:intro} 
% secluded
% multi-component
% indirect detection
% dark shower
Although the dark matter (DM) existence is confirmed by many astrophysical observations, the signs of DM direct detections and collider searches remain null, thus putting very stringent constraints on the coupling between the DM and standard model (SM) particles. 
Meanwhile, the Weakly-Interacting-Massive-Particle (WIMP) miracle for DM relic abundance can be also realized with DM evolving in a thermal bath of dark sector particles which have order one couplings and weak scale masses. 
The hidden sector interacts with the SM through suppressed portal interactions, evading the DM direct detection and collider searches. 
Such a scenario is dubbed secluded DM model~\cite{Pospelov:2007mp}. 
The existence of the hidden dark sector is well motivated by the fact that the DM is more abundant than {particles from SM which has complex particle spectra and gauge structure}, as well as theoretical perspectives such as string theory~\cite{Ibanez:1987sn,Blaszczyk:2014qoa}, Hidden Valley (HV) models~\cite{Strassler:2006im,Han:2007ae}, Dark QCD~\cite{Bai:2013xga} and so on~\cite{Escudero:2017yia,Barnes:2020vsc,Bringmann:2020mgx}. 
Generally, two or more particles in the dark sector could contribute to the measured DM density~\cite{Zurek:2008qg,Profumo:2009tb,Feldman:2010wy,Aoki:2012ub,Poulin:2018kap,Hall:2019rld,Hall:2021zsk}. 

Due to the large coupling required by the thermal relic density, the DMs around the {centers} of galaxies can annihilate into dark sector particles efficiently. 
The signature of this annihilation process provides the most robust probe to the secluded DM sector, although its manifestation is highly dependent on the model setup. 
In the simplest case, the DMs annihilate directly into the mediators which are interacting feebly with SM particles~\cite{Barnes:2021bsn}.   
{Then} the mediators can decay into SM particles, which induces stable photons, electrons/positrons, protons/antiprotons being detected in DM indirect detection experiments. 
In a non-trivial dark sector, the relic DMs may annihilate to other lighter dark states, with subsequent cascade decay~\cite{Elor:2015tva,Elor:2015bho,Beauchesne:2018myj,Beauchesne:2019ato,Kim:2019had}. A large number of mediator particles can be produced during the full evolution. 
Each additional dark sector particle in the cascade will increase the final state multiplicity, decrease the final state energy and broaden the final state spectra. 

In this paper, we consider another novel indirect detection signal for the secluded DM model in the case of heavy relic DM annihilating into boosted stable dark species with much smaller masses. 
If the light species {couple} to mediators with relatively large coupling~\cite{Hochberg:2014kqa} and the mediator is much lighter than the relic DM, the hierarchy between the energy and mass of the light species will induce copious radiations of the mediator after the annihilation. 
The phenomenon is known as dark shower~\cite{Cohen:2017pzm,Cohen:2020afv,Knapen:2021eip,Albouy:2022cin}. 
The boostness of light species opens up a new window to produce many kinds of light degrees of freedom in the dark sector. 
Probing the multiplicities and spectra of the radiated mediator can also help to reveal the inner structure of the dark sector. 
At the LHC, studies have shown that the dark jet from dark showering could be discriminated from QCD jet according to its substructures, such as semi-visible jet~\cite{Cohen:2015toa,Beauchesne:2017yhh}, emerging jet~\cite{Schwaller:2015gea}, {and} jet mass~\cite{Park:2017rfb}. 
The dark showers have also been studied in the context of indirect detection, for explaining the galactic center excess~\cite{Freytsis:2014sua,Freytsis:2016dgf,Curtin:2022oec}.
However, those studies assume the dark shower evolution to be QCD-like, {\it i.e.} under unbroken SU(N) gauge symmetry, where the mass effects in the splitting function are not fully taken into account. In particular, the radiation of the longitudinal component of the gauge boson is not considered.  
There {is} a number of works that study the mass effects in the dark matter shower in a simplified dark U(1) model framework with $Z^{\prime}$ mediator~\cite{Buschmann:2015awa,Kim:2016fdv,Chen:2018uii} and in the supersymmetric framework with scalar mediator~\cite{Li:2021bka}. 

This work will illustrate the DM indirect detection signal induced by the dark shower in the frameworks of two-component DM models with either vector or pseudoscalar mediator. 
There is a mass hierarchy between two DM particles and the heavier one contributes most of the DM relic~\cite{Planck:2015fie}. 
The dark shower is simulated by the Monte-Carlo method where the mass (symmetry breaking) effects are fully taken into account in the shower evolution. 
Assuming the vector and pseudoscalar mediators to be dominantly decaying into electron-positron pair and photon-pair respectively, we survey the constraints from Fermi-LAT observations of dwarf spheroidal galaxies~\cite{Fermi-LAT:2015att} and AMS02 measurement of positron flux~\cite{AMS:2021nhj,AMS:2019rhg}.

%\paragraph{Secluded dark matter models and their simplified scenarios}
\section{Secluded dark matter models and their simplified scenarios}

%In addition to the field content of the MSSM,  
As candidates of multi-DM models, 
we consider a dark/hidden sector that has dark/hidden local or global symmetry and some SM singlet fields which are charged under a dark/hidden symmetry.
Depending on the interactions between the hidden sector and the SM particles, typical scenarios of hidden sector models include the vector portal~\cite{Holdom:1985ag}, the (pseudo-)scalar portal~\cite{Silveira:1985rk,Patt:2006fw,McDonald:1993ex,DiazSaez:2021pmg}, and the neutrino portal~\cite{Minkowski:1977sc,Gell-Mann:1979vob,Yanagida:1979as,Mohapatra:1979ia}. In this study, we focus on our attention on the first two of these possibilities.

As simple UV complete models, we consider two models where one model provides vector portal while the other one induces pseudo-scalar portal interactions. \\
(i) {\it Vector portal model} : In this case we introduce a hidden local $U(1)_H$ symmetry and SM singlet field contents 
\begin{equation}
\text{Dirac fermions:} \ \chi \ (Q_\chi), \ \psi (Q_\psi), \quad \text{Scalar:} \ \varphi (2),
\end{equation} 
where the values inside {brackets} indicate $U(1)_H$ charges of the fields. 
The hidden gauge symmetry is assumed to be spontaneously broken by the {vacuum expectation value (VEV)} of $\varphi$, denoted by $v_\varphi$, and we have massive $Z'$ boson whose mass is $m_{Z'} = 2 g_H v_\varphi$.
In this study, the mixing between new scalar and the SM Higgs is taken to be small, so that the constraints from Higgs physics can be evaded.
The relevant Lagrangian for the Dirac fermions is 
\begin{equation}
\bar \chi (i \slashed{D} - m_\chi) \chi + \bar \psi (i \slashed{D} - m_\psi) \psi, 
\end{equation}
where $D_\mu \chi(\psi) = (\partial_\mu + i Q_{\chi(\psi)} g_H Z'_\mu) \chi(\psi)$ is the covariant derivative with $g_H$ being gauge coupling of $U(1)_H$.
In the following analysis, we parametrize gauge coupling with charge as {$g_{Z' \chi \chi  (Z' \psi \psi )} \equiv Q_{\chi(\psi)} g_H$.}
Here we require conditions $|Q_\chi| \neq |Q_\psi |$, $|2 Q_\chi| \neq 2$, $|2 Q_\psi| \neq 2$ and $| Q_\chi \pm Q_\psi | \neq 2$ so that we only have vector portal interaction, and
$\chi$ and $\psi$ do not mix. Then we have remnant discrete symmetry $Z_2^\chi \times Z_2^\psi$ where $\chi$ and $\psi$ are odd under each $Z_2$. Thus $\chi$ and $\psi$ are both stable. 
{Note also that we assume $Z'$ can decay into SM particles via small kinetic mixing effect inducing dark photon interactions.}\\
(ii) {\it Pseudo-scalar portal model}: In this case we introduce a hidden global $U(1)'_H \times Z^A_2 \times Z^B_2$ symmetry where $U(1)'_H$ is softly broken, and field contents are 
\begin{align}
& \text{{Fermions}}:  \ \chi_L (0, +, -), \ \chi_R (Q, +, -), \nonumber \\
&  \qquad \qquad \ \ \psi_L (0, -, +), \ \psi_R (Q, -, +), \nonumber \\
&  \text{Scalar :} \ \varphi' (-Q,+,+),
\end{align}
where {values inside brackets indicate} charges under $(U(1)'_H, Z^A_2, Z^B_2)$.
Fermions $\chi$ and $\psi$ can have Dirac masses $m_{\chi,\psi}$ since $U(1)'_H$ is assumed to be softly broken, and they are both stable due to $Z^A_2 \times Z^B_2$ symmetry.
We assume $\varphi'$ develops a VEV. 
Then {$\varphi'$} induces pseudo-Goldstone boson associated with spontaneous breaking of $U(1)'_H$ that has light mass $m_A$ due to soft $U(1)'_H$ breaking.
We thus have pseudo-scalar portal interactions
\begin{align}
\mathcal{L} \supset \ & y_\chi \overline{\chi_L} \chi_R \varphi' + y_\psi \overline{\psi_L} \psi_R \varphi' + h.c. \nonumber \\
& \supset i A ( y_{A \chi \chi}  \bar \chi \gamma_5 \chi + y_{A \psi \psi} \bar \psi \gamma_5 \psi),
\end{align} 
 where $\varphi' = (\phi + i A)/\sqrt{2}$ and $y_{A \chi \chi  (A \psi \psi )} \equiv y_{\chi(\psi)}/ \sqrt{2}$. We assume $\phi$ is much heavier than $A$ and pseudo-scalar portal interactions are dominant in DM annihilation processes.
{We assume $A$ decays into two photons via effective interaction of $A F_{\mu \nu} \tilde{F}^{\mu \nu}$ with $\tilde F^{\mu \nu} = \epsilon^{\mu \nu \alpha \beta} F_{\alpha \beta}$ that can be induced if there is a dark sector field interacting with both $A$ and photon.}
 
{In the above models DM is made of two WIMPs,  
and the interactions between the two {DM species} modify the Boltzmann equation and impact the computation of the relic density~\cite{Liu:2011aa,Belanger:2011ww}.
In the parameter region of interest, coupling constants are assumed to satisfy the relation $g_{Z' \psi \psi} > g_{Z' \chi \chi}$ ($y_{A \psi \psi} > y_{A \chi \chi}$) so that $\psi$ couples to $Z'(A)$ stronger than $\chi$.
In addition we assume $m_\chi > m_\psi$. 
Then relic density of $\chi$ is dominantly determined by annihilation cross section of $\bar \chi \chi \to Z'(A) \to \bar \psi \psi$ process since that of $\bar \chi \chi \to Z' Z'(AA)$ process
is more suppressed by smaller coupling $g_\chi (y_\chi)$.
The relic density in vector/pseudo-scalar {model is} thus roughly given by $(\Omega h^2)_{V(A)} \sim (0.1 {\rm pb})/\langle \sigma v\rangle_{\bar \chi \chi \to Z'(A) \to \bar \psi \psi} $.
Therefore relic densities can be estimated by model parameters such that
\begin{align}
\Omega h^{2}_{V} & \sim (0.05)  \left( \frac{1.0}{g_{Z' \psi \psi} } \right)^2  \left( \frac{0.01}{g_{Z' \chi \chi}}\right)^2  \left( \frac{m_\chi}{20~\text{GeV}}  \right)^2, \label{relicv} \\
\Omega h^{2}_{A} & \sim (0.1)  \left( \frac{1.0}{y_{A \psi \psi}} \right)^2  \left( \frac{0.01}{y_{ A \chi \chi}}\right)^2  \left( \frac{m_\chi}{20~\text{GeV}}  \right)^2, \label{relica}
\end{align}
where we assumed $m_{\psi, A/Z'}^2 \ll m_\chi^2$ and ignored $\psi$ and $Z'/A$ masses for simplicity.
When we set $m_{Z'/A} < m_\psi$ the relic density of $\psi$ will be much smaller than that of $\chi$ since $\psi$ couples to $Z'/A$ stronger than $\chi$ and 
$\psi$ {annihilates} into $Z'Z'(AA)$ efficiently.  
For a more accurate estimation of relic density, we apply two-component \texttt{micrOmegas}~\cite{Belanger:2014vza} implementing relevant interactions associated with DM candidates.}

%\paragraph{Showering of boosted dark matter}
\section{Showering of boosted dark matter}

From the annihilation of heavy relic DM $\chi$, boosted stable dark species including Dirac fermion $\psi$ and mediator (vector $Z^\prime$ or pseudoscalar $A$) are produced along with dark shower processes. 

For a collinear time-like branching process $a\to b+c$, where the off-shell particle $a$ is in the final state of a preceding hard process, we parameterize the four-momentum of these particles by
\begin{align}
&P_a^\mu=(P,~0,~0,~P-\frac{k_{\perp}^2+\bar{z}m_b^2+zm_c^2}{2z\bar{z}P})~,\\
&P_b^\mu=(zP,~k_{\perp},~0,~zP-\frac{k_{\perp}^2+m_b^2}{2zP})~,\\
&P_c^\mu=(\bar{z}P,~-k_{\perp},~0,~\bar{z}P-\frac{k_{\perp}^2+m_c^2}{2\bar{z}P})~,
\end{align}
where $\bar{z}=1-z$, $z$ ranges in $(0,~1)$, and $P^2\gg k_{\perp}^2,m_i^2~(i=a,b,c)$. When ignoring terms proportional to $\frac{k_{\perp}^2~\text{or}~m_i^2}{P^2}~(i=a,b,c)$, $b$ and $c$ are on-shell, but $a$ is off-shell with virtuality $Q$ satisfying
\begin{align}
Q^2=\frac{k_{\perp}^2+\bar{z}m_b^2+zm_c^2}{z\bar{z}}~.
\end{align}
The differential cross section of the hard process followed by the branching of $a\to b+c$ can be expressed as   
\begin{equation}
d \sigma_{X, b c} \simeq d \sigma_{X, a} \times d \mathcal{P}_{a \rightarrow b+c}~,
\end{equation}
where $X$ stands for other particles in the final state of the hard process besides $a$. And $d \mathcal{P}_{a \rightarrow b+c}$ is the differential splitting function for the $a \to b+c$ branching, which can be expressed as 
\begin{equation}
\frac{d \mathcal{P}_{a \rightarrow b+c}}{d z d \ln Q^2} \approx \frac{1}{N} \frac{1}{16 \pi^2} \frac{Q^2}{\left(Q^2-m_a^2\right)^2} {\left|M_{\text {split }}\right|^2}~,
\end{equation}
where $N=2$ if $b$ and $c$ are identical particles otherwise $N=1$. And ${\left|M_{\text {split }}\right|^2}$ is the matrix-element square of the $a\to b+c$ branching process by considering the amputated $a\to b+c$ Feynman diagram with on-shell particle polarizations.

With two different types of mediators (vector boson $Z^\prime$ and pseudoscalar $A$), the $\left|M_{\text{split}}\right|^2$ of branchings from $\psi$ and mediator is listed in Tab.~\ref{tab:splitff} and \ref{tab:splitf}, respectively. We have to emphasize that the $\left|M_{\text{split}}\right|^2$ in these tables have been averaged (summed) over polarizations of the corresponding initial (final) particles within the requirement of fermion helicity relation given in the second column. And according to the Ref.~\cite{Chen:2016wkt}, we have eliminated terms proportional to $(Q^2-m_a^2)$ in $M_{\text{split}}$ of processes involving the longitudinal mode of $Z^\prime$.

\begin{table}[htbp]
	\centering
	\scalebox{0.82}{
	\begin{tabular}{ccc}  
		\hline
		Process &  $\lambda_a(\lambda_b),~\lambda_c$ & ${\left| M_{\text{split}}\right|^2}$ \\
		\hline 
		$Z_{\text{T}}^\prime \to \psi +\bar{\psi}$ &  $\lambda_b=\lambda_c$ & $2 {g^2_{Z^\prime \psi\psi} }\frac{m_\psi^2}{z(1-z)} $ \\
		$Z_{\text{T}}^\prime \to \psi +\bar{\psi}$ &  $\lambda_b=-\lambda_c$ & $2 {g^2_{Z^\prime \psi\psi} } \big(z^2+(1-z)^2 \big)\big(Q^2-\frac{m_\psi^2}{z(1-z)} \big) $ \\
		$Z_{\text{L}}^\prime \to \psi +\bar{\psi}$ &  $\lambda_b=\lambda_c$ & 0 \\
		$Z_{\text{L}}^\prime \to \psi +\bar{\psi}$ &  $\lambda_b=-\lambda_c$ & $8 {g^2_{Z^\prime \psi\psi } } m_{Z^\prime}^2 z(1-z)$ \\
		$\psi/\bar{\psi} \to Z_{\text{T}}^\prime +\psi/\bar{\psi}$ & $\lambda_a=\lambda_c$ & 
		$2 {g^2_{Z^\prime \psi\psi } }  \frac{\big(1+(1-z)^2\big)}{z}\big(Q^2-\frac{\big(m_{Z^\prime}^2(1-z)+m_{\psi}^2z\big)}{z(1-z)} \big)$ \\
		$\psi/\bar{\psi} \to Z_{\text{T}}^\prime +\psi/\bar{\psi}$ & $\lambda_a=-\lambda_c$ & 
		$2 {g^2_{Z^\prime \psi\psi }} \frac{m_\psi^2 z^2}{1-z}$ \\
		$\psi/\bar{\psi} \to Z_{\text{L}}^\prime +\psi/\bar{\psi}$ & $\lambda_a=\lambda_c$ & 
		$4 {g^2_{Z^\prime \psi\psi } }\frac{m_{Z^\prime}^2(1-z)}{z^2}$ \\
		$\psi/\bar{\psi} \to Z_{\text{L}}^\prime +\psi/\bar{\psi}$ & $\lambda_a=-\lambda_c$ & 0 \\
		\hline
	\end{tabular}}
	\caption{\label{tab:splitff} The $\left| M_{\text{split}}\right|^2$ of branchings from $\psi$ and mediator when the mediator is vector boson $Z^\prime$. $Z^\prime_{\text{T}}$ and $Z^\prime_{\text{L}}$ are the transverse and longitudinal polarization modes of $Z^\prime$, respectively. The fermion helicity is labelled by $\lambda$.}
\end{table}

\begin{table}[htb]
	\centering
	\begin{tabular}{ccc}  
		\hline
		Process &  $\lambda_a(\lambda_b),~\lambda_c$ & ${\left| M_{\text{split}}\right|^2}$ \\
		\hline 
		$A \to \psi +\bar{\psi}$ &  $\lambda_b=\lambda_c$ & $2 {y^2_{A\psi\psi } } \big(Q^2-\frac{m_\psi^2}{z(1-z)} \big) $ \\
		$A \to \psi +\bar{\psi}$ &  $\lambda_b=-\lambda_c$ & 
		$2 {y^2_{A \psi\psi } }  \frac{m_\psi^2}{z(1-z)}$ \\
		$\psi/\bar{\psi} \to A +\psi/\bar{\psi}$ & $\lambda_a=\lambda_c$ & 
		$ {y^2_{A \psi\psi } }  \frac{m_\psi^2 z^2}{1-z}$ \\
		$\psi/\bar{\psi} \to A +\psi/\bar{\psi}$ & $\lambda_a=-\lambda_c$ & 
		$ {y^2_{A \psi\psi }}\big( Q^2z-\frac{m_\psi^2 z+m_A^2(1-z)}{1-z} \big) $ \\
		\hline
	\end{tabular}
	\caption{\label{tab:splitf} The $\left| M_{\text{split}}\right|^2$ of branchings from $\psi$ and mediator when the mediator is pseudoscalar $A$. The fermion helicity is labelled by $\lambda$.}
\end{table}

Considering all the possible branching of $a$, the famous Sudakov form factor 
\begin{align}
&\Delta_a\left(Q_{\max } ; Q_{\min }\right)=\nonumber\\
&\text{exp}\left[-\sum_{b c} \int_{\ln Q_{\min}^2}^{\ln Q_{\max }^2} d \ln Q^2  \int_{z_{\min }(Q)}^{z_{\max }(Q)} d z\frac{d \mathcal{P}_{a \rightarrow b+c}(z, Q)}{d z d \ln Q^2}\right]~,
\end{align} 
gives $a$'s probability of evolving from $Q_{\max}$ to $Q_{\min}$ with no branching. The allowed $z$ range at $Q$ is given by~\cite{Li:2021bka}
\begin{align}
	z_{\min }(Q) & =\frac{1-v_c}{1+v_c / v_b}~, \\
	z_{\max }(Q) & =\frac{1+v_c}{1+v_c / v_b}~,
\end{align}
with
\begin{align}
	v_b & =\sqrt{1-\left(\frac{2 m_b Q}{Q^2+m_b^2-m_c^2}\right)^2}~, \\
	v_c & =\sqrt{1-\left(\frac{2 m_c Q}{Q^2+m_c^2-m_b^2}\right)^2}~.
\end{align}

A numerical Monte Carlo method with Markov chain based on the Sudakov factors of $\psi$ and mediator ($Z^\prime$ or $A$) is used to study the dark parton shower in this work. When evolving from a high virtuality scale $Q_{\text{max}}$, chosen to be the CM-frame energy of the hard annihilation process of $\chi$, down to a low scale $Q_{\text{min}}$ with small $Q$ steps, if the $a\to b+c$ branching occurs at some $Q$, the evolution will be carried on with both the daughters $b$ and $c$.

%\paragraph{Indirect detection signals}
\section{Indirect detection signals}

\begin{figure}[htb]
\centering
\includegraphics[width=0.2\textwidth]{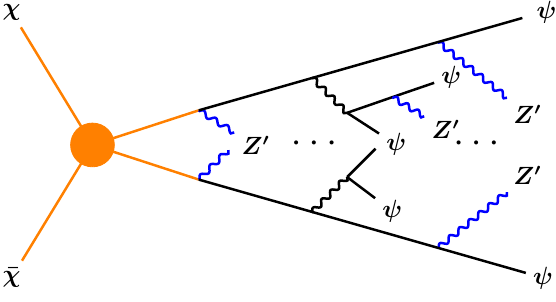}
\includegraphics[width=0.17\textwidth]{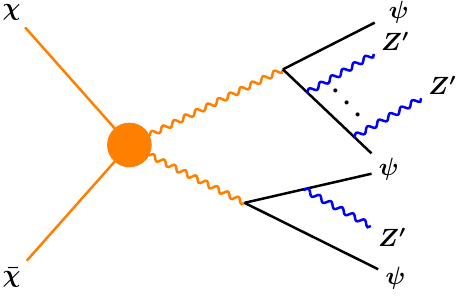}
\includegraphics[width=0.1\textwidth]{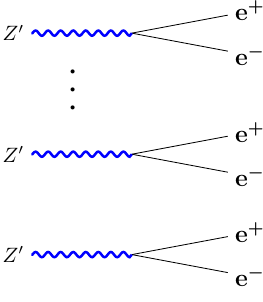}
\caption{Indirect detection signature for $\chi \chi$ annihilation in the  vector portal model. 
\label{feynman}}
\end{figure}

Considering the dark shower, {the annihilation of relic DM ($\chi$)} for the vector portal model will induce multiple $Z^{\prime}$ {emissions} as illustrated in Fig.~\ref{feynman}. Similarly, there will be multiple pseudoscalar signal for the pseudoscalar portal model, except that the $\chi \chi \to A A $ annihilation is $p$-wave suppressed.  

The dwarf spheroidal satellite galaxies (dSphs) of the Milky Way contain a substantial DM component\cite{Mateo:1998wg,McConnachie_2012} and are expected to produce some of the brightest signals of DM annihilation, thus they can be used to set stringent limits on {the pseudoscalar portal DM model}. 
%Compared to signals from the galactic center, the dSphs also have a huge reduction in the astrophysical background. 
%The gamma rays from dwarfs observed by the
%Fermi Large Area Telescope (LAT) is used to set stringent limits on DM models, which have a high sensitivity to energies ranging from 20 MeV to $\gtrsim$ 300 GeV. 
We use the publicly released bin-by-bin likelihoods of each dwarf in Ref.~\cite{Fermi-LAT2015att}. Treating the energy bins as independent, the multiplication of likelihoods of all of the bins gives the likelihood for a given dwarf $i$, $\mathcal{L}_{i} ( \Phi_{\gamma} | \mathcal{D}_{i})$, in which $\Phi_{\gamma}$ is the gamma-ray flux of the DM model and  $\mathcal{D}_{i}$ is the data. 
Finally, the full likelihood is obtained by multiplying the likelihood of the following 15 dwarfs: Bootes I, Canes Venatici II, Carina, Coma Berenices, Draco, Fornax, Hercules, Leo II, Leo IV, Sculptor, Segue 1, Sextans, Ursa Major II, Ursa Minor, and Willman 1.

Giving the energy spectra of gamma-rays per annihilation of heavier {DM ($\chi$) } as discussed above, the differential flux of gamma-rays at the location of the earth is given by~\cite{Cirelli:2010xx}
\begin{equation}
 \frac{d\Phi_\gamma}{dE_\gamma}=\frac{1}{\eta}\frac{1}{4\pi}\frac{1}{m_\chi^2}J \sum_{f=\psi,A}\left \langle\sigma_A v\right\rangle _f\frac{dN^f_\gamma}{dE_\gamma}  \label{fluxa}~,
\end{equation}
where $\eta$ is 4 for Dirac DM and 2 for Majorana DM.  The $J$-factor is the line-of-sight
(l.o.s.) integral through the DM distribution integrated over a solid angle. 
We adopt the value of $J$-factors for the Milky Way dSphs in Ref.~\cite{Fermi-LAT2015att}, which are calculated assuming an Navarro-Frenk-White density profile and
integrated over a circular region with a solid angle of $\Delta \Omega  \sim 2.4\times 10^{-4}$ sr. The $dN^f_\gamma / dE_\gamma$ is the energy spectrum of gamma-rays per annihilation in the channel with final state $f$.
The thermal averaged annihilation cross section $\left \langle\sigma_A v\right\rangle _A$ (measured in $\mathrm{cm}^3 ~\mathrm{s}^{-1}$) of the channel $\chi\chi\rightarrow A A$ is $p$-wave suppressed, {while} the cross section for $s$-wave annihilation $\chi\chi\rightarrow \psi\psi$ is given by
\begin{equation}
\left \langle\sigma_A v\right\rangle _\psi= {y_{A \chi\chi }^2 y_{A \psi\psi }^2 }  \frac{m_\chi\sqrt{m_\chi^2-m_\psi^2}}{2\pi(m_A^2-4m_\chi^2)^2}~. 
\end{equation}

% ams
The AMS measurement of {positron} flux~\cite{PhysRevLett.113.121102,PhysRevLett.113.121101,AMS:2021nhj} could set stringent limits to the vector portal model.
%Apart from the signal of gamma-rays, the precise measurements of positrons in the energy range of $\sim$ 1GeV  to $\sim$ 500GeV has been released recently by AMS-02\cite{PhysRevLett.113.121102,PhysRevLett.113.121101}. The reported flux of positrons is used to set stringent limits on general DM annihilation scenarios\cite{AGUILAR20211}.
In contrast to the gamma-ray, positrons 
propagate through the galactic magnetic field are deflected by its irregularities, which {need} to be investigated numerically. 
Refs.~\cite{Cirelli:2010xx,TP-toolbox-web} provide essential propagation functions that encode all the intervening astrophysics. Thus, the differential flux at the {location} of the earth can be calculated by convoluting the spectra at production with the propagation functions: 
\begin{equation} \label{fluxpos}
\begin{split}
&\frac{d\Phi_{e^+}}{dE_{e^+}}(E)=\frac{v_{e^+}}{4\pi b(E,r_{\rm sun})} \frac{1}{\eta}\left(\frac{\rho(r_{\rm sun})}{{m_{\chi}}}\right)^2 
\\
&\times\sum_{f=\psi,Z'}\langle\sigma v\rangle_f \int_E^{m_{\chi}} d E_{\mathrm{s}} \frac{d N_{e^{+}}^f}{d E}\left(E_{\mathrm{s}}\right) I\left(E, E_{\mathrm{s}}, r_{\rm sun}\right) \, .
\end{split}
\end{equation}
The information on the galactic DM density profile and propagation of positrons is summarized in the halo function $I\left(E, E_{\mathrm{s}}, r_{\text{sun}}\right)$, where $E_{\mathrm{s}}$ is the energy at production and $r_{\rm sun}$ is the Earth distance to the galactic center.
The energy loss coefficient function $b(E,r_{\rm sun})$  depicts the energy loss at the location of the earth due
to several processes, such as synchrotron radiation and Inverse Compton scattering (ICS) on CMB photons and on infrared or optical galactic starlight. 
We adopt the MED model~\cite{Delahaye_2008,Donato_2004} for the propagation parameters and {MF1}~\cite{Buch:2015iya} for the magnetic field configuration. Different choices of parameters can affect the flux up to one order of magnitude. 
Moreover, $r_s=24.42 ~\mathrm{kpc}$ and $\rho_s=0.184~\mathrm{GeV/cm^3}$ are adopted in the Navarro-Frenk-White density profile.  
The thermal averaged annihilation cross sections for the vector portal model are given by
\begin{equation}
    \langle\sigma v\rangle_f=\left\{\begin{array}{ll}
   {g_{Z' \chi\chi }^4 } \frac{(m_\chi^2-m_{Z'}^2)^{3/2}}{4\pi m_\chi(m_{Z'}^2-2m_\chi^2)^2} & \chi\chi\rightarrow Z' Z' \\
    {g_{Z' \chi\chi }^2 g_{Z' \psi\psi}^2 } \frac{\sqrt{m_\chi^2-m_\psi^2}(2m_\chi^2+m_\psi^2)}{2\pi m_\chi(m_{Z'}^2-4m_\chi^2)^2} & \chi\chi\rightarrow \psi\psi 
\end{array}\right. ~.
\end{equation}

% ams data and bound
In order to derive the upper {limits} on the cross sections or couplings, we assume the positron flux from AMS-02 measurement arises solely from the astrophysical backgrounds and fit the $\log (\Phi_{e^{+}})$ of background with degree 6 polynomial of $\log E_{e^+}$.
Defining $\chi^{2} = \sum_{i} \frac{\Phi_{i}^{\text{model}} (\alpha)- \Phi_{i}^{\text{data}}}{\sigma^{2}_{i}}$, where  $\Phi_{i}^{\text{model}}$, $\Phi_{i}^{\text{data}}$ and $\sigma_{i}$ represent the flux predicted by the polynomial function with parameters $\{ \alpha \}$, the measured flux and the total uncertainties (systematic and statistical uncertainties added in quadrature) in the $i$-th energy bin respectively. {And} the best-fit values for the polynomial parameter and $\chi^{2}$ are denoted by $\{ \alpha_{\text{bf}}\}$ and $\chi^{2}_{\text{bf}}$. 
Then we add the DM-induced flux in Eq.~(\ref{fluxpos}) to the background and fit the stacked flux allowing the parameters to float within 30\% of $\{ \alpha_{\text{bf}}\}$. 
The 95\% C.L. limit can be obtained by $\chi^{2} (\langle \sigma v \rangle_{95}) \equiv \chi^{2}_{\text{bf}} +2.71$. This methodology has been widely used in literature~\cite{Elor:2015bho,Leane:2018kjk,Dutta:2022wuc}. 

% end

%\paragraph{Constraints and discussion}

\section{Constraints and discussion}

The above methodology has been applied to the full parameter space of our models. 
The details for some benchmark points are provided in the supplemental material. 
In Fig.~\ref{fig.bounds}, we show the bounds from AMS-02 (shaded region) and Fermi-LAT observations of dwarf galaxies (solid line) in  $m_\psi$-$g_\psi$ plane with various $m_\chi s$ {where}
{$g_\psi = g_{Z' \psi \psi}~(y_{A \psi \psi}$) for vector (pseudoscalar) mediator scenario. }

\begin{figure}[htb]
\centering
\includegraphics[width=0.5\textwidth]{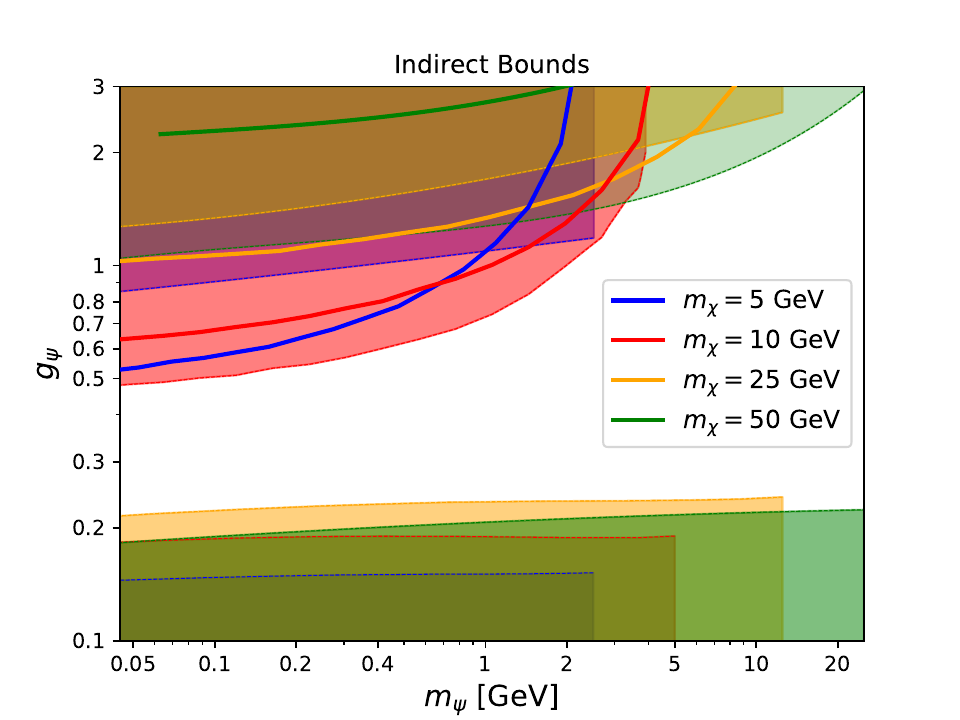}
\caption{Indirect detection bounds from Fermi-LAT for pseudoscalar portal model (solid line) and from AMS-02 for vector portal model (shaded region). 
\label{fig.bounds}}
\end{figure}

The Fermi-LAT bound and the upper exclusion region of AMS-02 are induced by the dark showers subsequent to the annihilation of $\chi \chi \to \psi \psi$. 
In the pseudoscalar portal model, giving the correct DM relic density ({the parameter relation} {approximately} satisfies Eq.~(\ref{relica})), the gamma-ray flux (in Eq.~(\ref{fluxa})) is proportional to  $g_\psi^2 / m_\chi^2 $ in the region  $m_{\psi, A}\ll m_\chi$. 
As a result, the sensitivity of Fermi-LAT degrades with increasing $m_{\chi}$ in the small $m_\psi$ region. 
In the vector portal model, the positron flux from $\chi \chi \to \psi \psi$ annihilation is proportional to $g^2_\psi/m_\chi^2$ {which is} similar to the gamma-ray flux of the pseudoscalar model (more details are provided in supplementary). 
The bounds in the small $m_\psi$ region fluctuate with increasing $m_{\chi}$ mainly attributed to the features of the AMS-02 data and the $\chi^{2}$ test method. 
In both cases, the larger mass splitting between the $\chi$ and $\psi$ can lead to stronger dark showering effects,  {\it i.e.} higher particle multiplicity. Thus the sensitivities become stronger with decreasing $m_\psi$ for a given $m_{\chi}$. 
The additional $s$-wave annihilation $\chi \chi \to Z^{\prime} Z^{\prime}$ in the vector portal model induces the lower exclusion region of AMS-02, which is the only detectable region if dark showering effects are ignored. 
The flux of this channel is proportional to $1/g^4_\psi$ in the region with $m_{\psi, Z'} \ll m_{\chi}$ and $g_{Z' \psi \psi} \lesssim \mathcal{O}(1)$ so that the bounds are relatively stable with respect to the variation of $m_{\chi}$.

%Assuming the $Z^{\prime} \to e^{+}e^{-}$ or $\phi \to \gamma \gamma$,  \\
%polatization information for  $Z^{\prime}$ being taken into account\\
%Consider the AMS positron data and  Dwarf Spheroidal Galaxies gamma ray data\\
%The flux of gamma ray\\
%The flux of positron including the propagation effects\\

%constraints for benchmark points \\
%chi square - likelihood\\
%discuss about the variation that either increase or release the bounds \\

%without parton shower, only the lower bounds exsit. 

\begin{acknowledgments}
This work was supported in part by the National Natural Science Foundation of China under grants No. 11905149 and No.12247119, by the Natural Science Foundation of Sichuan Province under grants No. 2023NSFSC1329, by the Fundamental Research Funds for the Central Universities. C.Z. acknowledges the Sino-German (CSC-DAAD) Postdoc Scholarship Program, 2023 (57678375).
\end{acknowledgments}

\bibliography{multiDM}

%%%%%%%%%%%%%%%%%%%%%%%%%%%%%%%%%%%%%%%%%%%%%%%%%%%%%%%%%%%%%%%%%%%%

\clearpage

\begin{widetext}
\section*{Appendices}

\subsection{Benchmark points and dark shower spectra}
\label{supp:bps}
To illustrate the dark showering effects for the DM annihilation in the pseudoscalar portal model and vector portal model, we select 5 benchmark points in each model and provide their details in Tab.~\ref{tab:bps}. 
The couplings $g_{Z^{\prime}\chi\chi}$ and ${y_{A \chi\chi}}$ have been appropriately tuned to guarantee the relic density $\Omega h^{2}_{\chi} \sim 0.12$ and $\Omega h^{2}_{\psi} \ll \Omega h^{2}_{\chi}$. 
Note that in calculating the bounds in Fig.~\ref{fig.bounds}, the couplings $g_{Z^{\prime}\chi\chi}$ and ${y_{A \chi\chi}}$ are determined in more refined ways (logarithmically {scanning} the couplings $g_{Z^{\prime}\chi\chi}$ and ${y_{A \chi\chi}}$ from $10^{-3}$ with ratio $10^{0.001}$ until the total density of $\chi$ and $\psi$ reaches 0.12). 
The $\chi^{2}$ test value for each benchmark point of the vector portal model using the AMS-02 data, as well as the Fermi-LAT limit and the theoretical cross section for each benchmark point of the pseudoscalar portal model {is} provided. 

\begin{table}[htb]
	\centering
	\scalebox{1.0}{
	\begin{tabular}{|c|c|c|c|c|c|} \hline
	 & A & B & C & D & E\\ \hline
  $m_{\chi}$ [GeV] & 100  & 10 & 1000 &10 & 1000 \\\hline
  $m_{\psi}$ [GeV] & 1 & 1 & 1 & 0.1 & 10 \\ \hline
  $g_{Z^{\prime}\chi\chi}$ & 0.029 &  0.003 &  0.3 & 0.003 & 0.3 \\ \hline
  ${y_{A \chi\chi} }$ & 0.02 & 0.0022 & 0.22 & 0.0022 & 0.22 \\ \hline
  $\Omega_{\chi}$ [Vector]  & $0.111$ & $0.115$ & $0.101$ & 0.115 & 0.101 \\ \hline
  $\Omega_{\chi}$ [Pseudoscalar]  &  $0.128$ & $0.118$  & $0.108$  & $0.117$  & $0.108$  \\ \hline
  $\Omega_{\psi}$ [Vector]& $2.03 \times 10^{-7}$ &  $2.17 \times 10^{-6}$ & $2.24 \times 10^{-8}$ & $2.17\times 10^{-8}$ & $1.93 \times 10^{-6}$ \\ \hline
  $\Omega_{\psi}$ [Pseudoscalar] &  $1.54\times 10^{-6}$ & $2.74\times 10^{-5}$  & $3.07\times 10^{-7}$   & $1.21\times 10^{-7}$  & $1.39 \times 10^{-5}$ \\ \hline
  AMS-02 $\chi^{2}$ &  62.12  & 2001.9 & 50.0 & 2822.2 & 49.9 \\ \hline
  Fermi-LAT limit [cm$^{2}$]   & $9.07533 \times 10^{-26}$  & $5.28983 \times 10^{-27}$ &  $2.47742\times 10^{-24}$ & $2.50722\times 10^{-27}$ & $3.38713 \times 10^{-24}$ \\ \hline
  Model cross section [cm$^{2}$]  & $4.18391\times 10^{-26}$ & $5.04364 \times 10^{-26}$ & $5.06272 \times 10^{-26}$ & $5.06253\times 10^{-26}$ & $5.06253 \times 10^{-26}$ \\  \hline
	\end{tabular}}
	\caption{\label{tab:bps} Benchmark points for vector portal DM model and pseudoscalar DM model. In vector portal case, the coupling of $\psi$ is chosen as $\alpha_{D} \equiv {g_{Z'\psi \psi}^{2}}/ 4\pi =0.2$ and the $\chi^{2}_{\text{bf}}$=49.8. In the pseudoscalar case, the coupling of $\psi$ is ${y_{A \psi \psi} }=3$. Varying the masses of mediators ($Z^{\prime}$ and $A$) in the range $\mathcal{O}(0.1)m_{\psi}$ gives similar results. In the last and second to the last row, the theoretical value and the Fermi-LAT bound of the $\chi\chi \to \psi \psi$ annihilation cross section in the pseudoscalar portal model are presented. } 
\end{table}

In Fig.~\ref{fig:spec}, we plot the spectra of positron and gamma-ray after the dark shower and mediator decay for the annihilation channels $\chi \chi \to \psi \psi$ and $\chi \chi \to Z^{\prime} Z^{\prime}$ in the vector portal model, $\chi \chi \to \psi \psi$ in the pseudoscalar portal model. 
In the decay of the $Z^{\prime}$, the polarization information has been taken into account. 
Comparing the left and right panels of the figure, we can find that the dark shower is copious for $\psi$, and it produces a harder spectrum in the pseudoscalar model than in the vector model. 
{However,} the shower of the boosted $Z^{\prime}$ in the $\chi \chi \to Z^{\prime} Z^{\prime}$ channel is rare, giving the peak of {$E_{Z^{\prime}}$} distribution at around $m_{\chi}$.  
The heights of the peaks are close to 2 because there are two $Z^{\prime}$ produced in each annihilation. 

\begin{figure}[htb]
\centering
\includegraphics[width=0.32\textwidth]{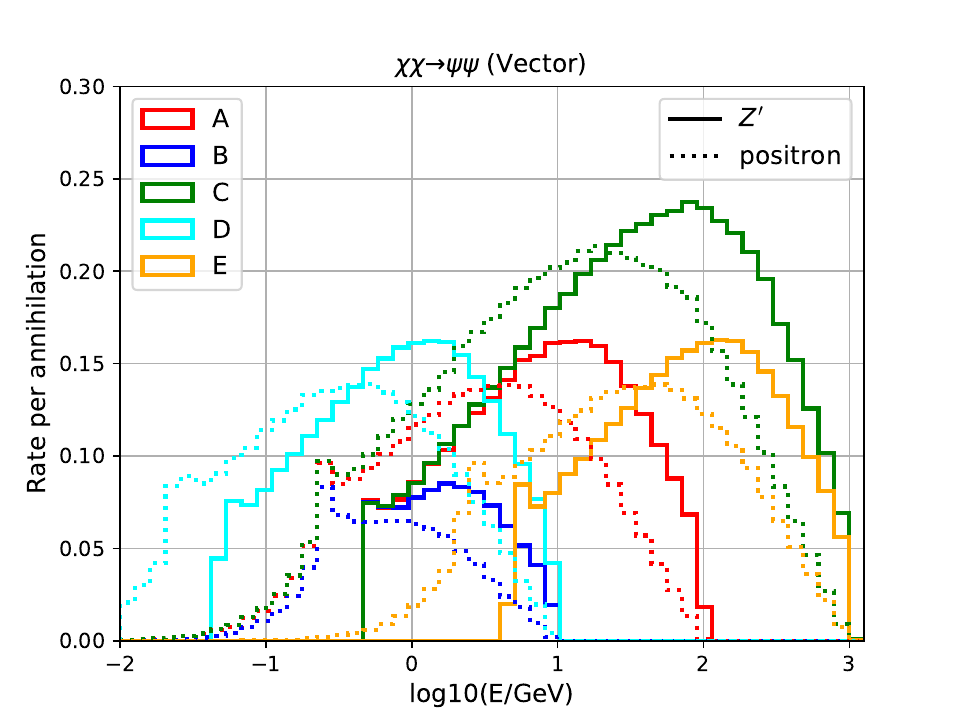}
\includegraphics[width=0.32\textwidth]{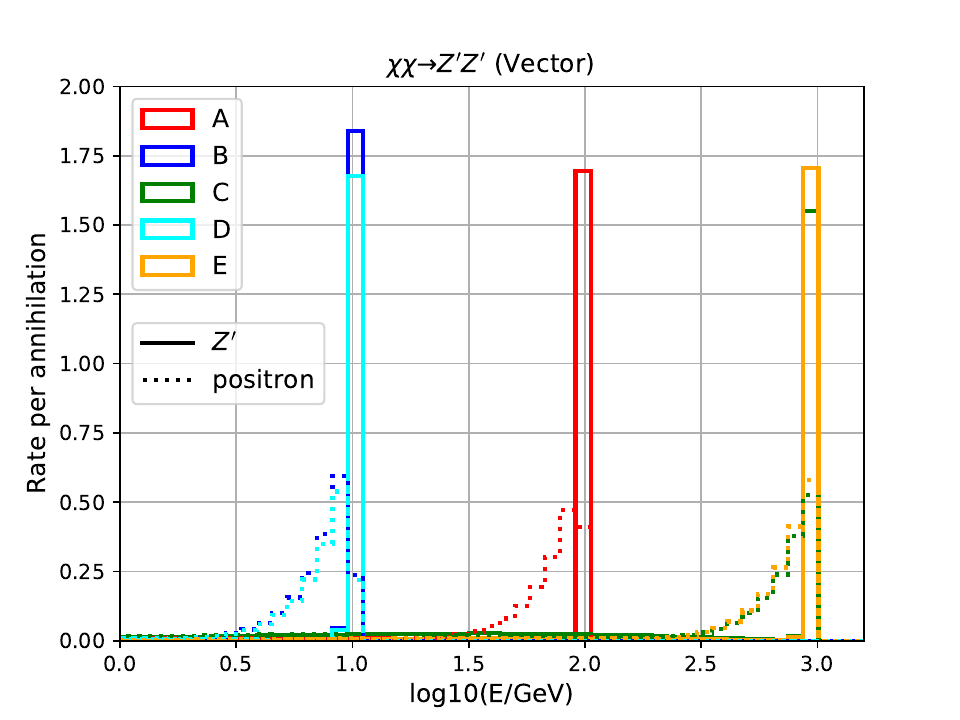}
\includegraphics[width=0.32\textwidth]{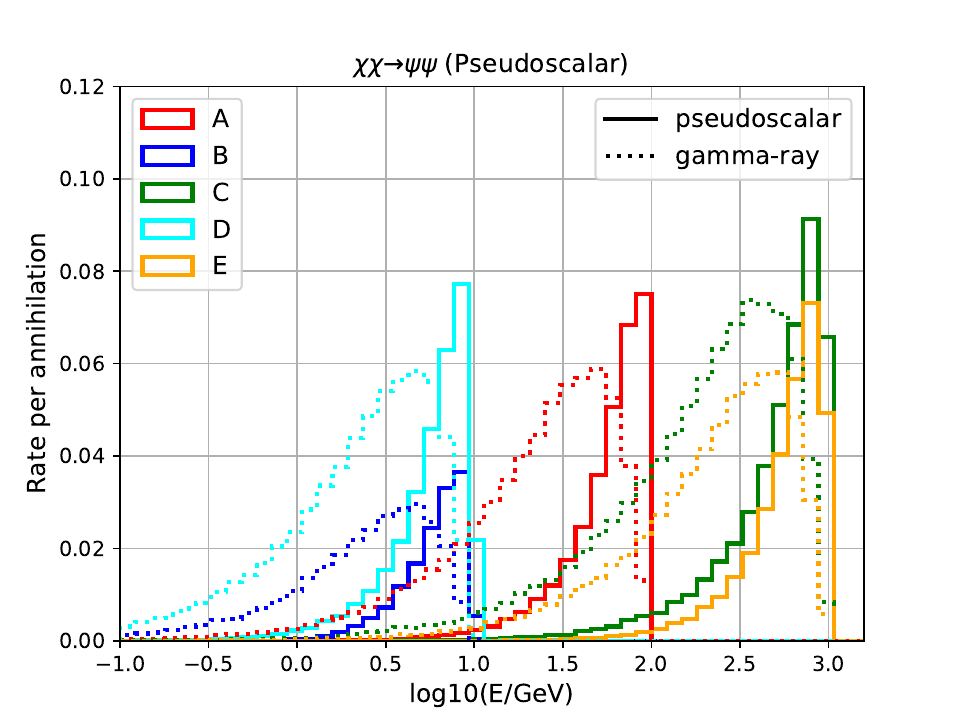}
\caption{Shower spectra for benchmark points. The mediator ($Z^{\prime}$ and pseudoscalar) masses are chosen to be half of the light DM $\psi$ mass. 
\label{fig:spec}}
\end{figure}

\subsection{Features of the positron flux}
\label{supp:flux}

\begin{figure}[htb]
    \centering
    \includegraphics[width=0.48\textwidth]{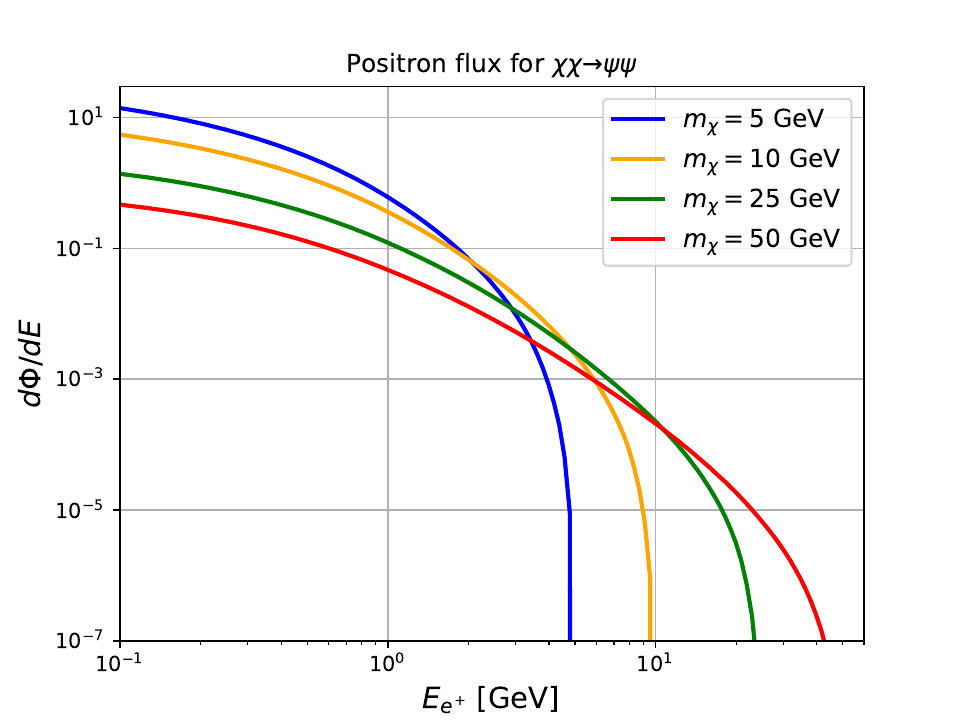}
    \includegraphics[width=0.48\textwidth]{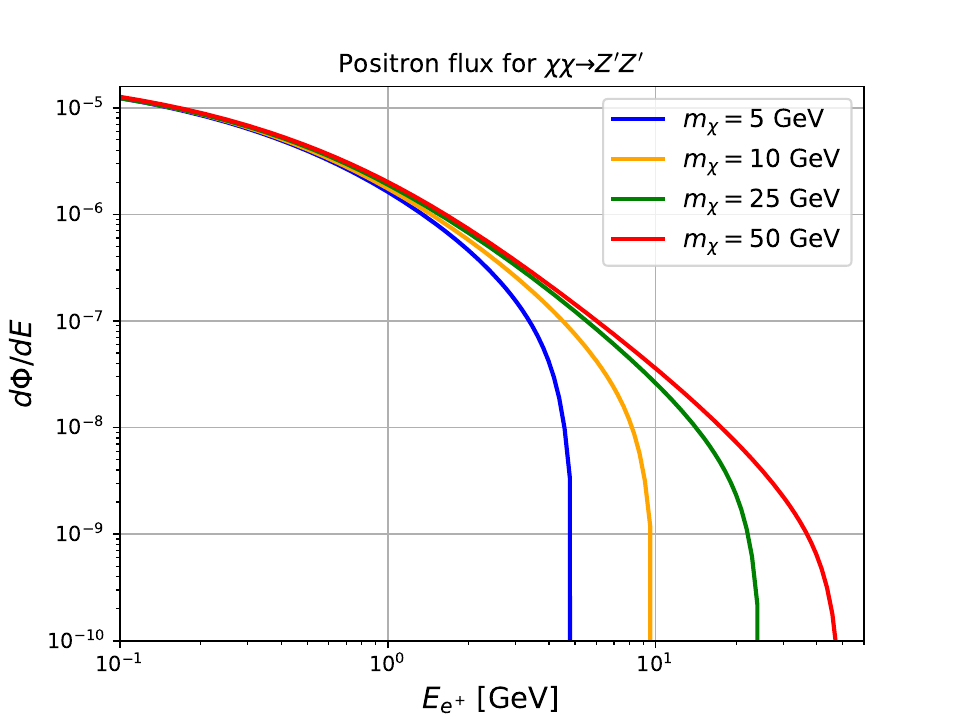}
    \caption{Fluxes of positron near earth for the $\chi\chi\rightarrow\psi\psi$ channel (left panel) and $\chi\chi\rightarrow Z'Z'$ channel (right panel). The following parameters are chosen: $m_\psi=0.119~\mathrm{GeV}$, $m_Z'=0.5 m_\psi$, ${g_{Z'\psi \psi}}=\sqrt{4\pi \times 0.2}$ and {$g_{Z' \chi \chi}$} is fixed by Eq.~(\ref{relicv}). \label{fig:fluxes}}
    \end{figure}

We discuss the features of the positron flux for the vector portal model in this section. The corresponding results for the gamma-ray flux in the pseudoscalar portal model are similar to {those} of the $\chi \chi \to \psi \psi$ channel in the vector portal model. 

In the limit of $m_\chi\gg m_{\psi,Z'}$ and assuming the correct relic density (with parameter relation satisfies Eq.~(\ref{relicv})), the positron {fluxes} of $\chi \chi \to \psi \psi$ channel ($\frac{d \Phi}{d E}\Big{|}_{\psi}$) and $\chi \chi \to Z^{\prime} Z^{\prime}$ channel ($\frac{d \Phi}{d E}\Big{|}_{Z'}$) can be written as 
\begin{equation}
    \begin{aligned}
        &\frac{d \Phi}{d E}\Big{|}_{\psi}= \frac{C_1(E)C_{2,Z'}}{m_\chi^2} \int_E^{m_\chi} d E_{\mathrm{s}} \frac{d N^\psi_{e^{+}}}{d E}\left(E_{\mathrm{s}}\right) I\left(E, E_{\mathrm{s}}, r_{\text{sun}}\right)~,
        \\
        &\frac{d \Phi}{d E}\Big{|}_{Z'}= \frac{C_1(E)C_{2,{Z'}}^2}{g_\psi^4} \int_E^{m_\chi} d E_{\mathrm{s}} \frac{d {N}^{Z'}_{e^{+}}}{d E}\left(E_{\mathrm{s}}\right) I\left(E, E_{\mathrm{s}}, r_{\text{sun}}\right)~,
        \end{aligned}
\end{equation}
{where $d N_{e^+}^{\psi} /d E ~(\propto \alpha_{D})$ and $d {N}^{Z'}_{e^{+}} / d E~(\propto [ 1 + \frac{\alpha_{D}}{0.2} (N_{D}-1)]$, $N_{D}$ is the total number of simulated positrons for $\alpha_{D}=0.2$) are the spectra of the positron at the source for the $\chi \chi \to \psi \psi$ channel and the $\chi \chi \to Z' Z'$ channel respectively. 
For a given $E$, the halo function $I\left(E, E_{\mathrm{s}}, r_{\text{sun}}\right)$ is relatively flat with respect to the energy at the source $E_{s}$. 
The factor $C_{2,{Z'}}$ corresponds to the constant obtained with Eq.~(\ref{relicv}), {\it i.e.} $C_{2,{Z'}}= g_{Z' \psi \psi}^2 g_{Z' \chi \chi}^2 / m_\chi^2 $. 
The factor $C_1(E)$ collects the rest astrophysical parameters in Eq.~(\ref{fluxpos}). }

Fig.~\ref{fig:fluxes} shows the positron fluxes around the earth for the  $\chi\chi\rightarrow\psi\psi$ channel and the $\chi\chi\rightarrow Z'Z'$ channel with several given $m_{\chi}$. 
The flux for the $\chi\chi\rightarrow\psi\psi$ channel is proportional to $1/m^{2}_{\chi}$ thus is suppressed for heavier $\chi$. Although the dark shower becomes more copious for larger $m_{\chi}$, {its effect is subdominant compared to that of the factor $1/m^{2}_{\chi}$}. 
The difference in the positron flux of the $\chi\chi\rightarrow Z'Z'$ channel for different $\chi$ mass is attributed to the dark shower process. Heavier $\chi$ will give rise to higher positron flux due to a longer evolution period. 
It should be noted that the relations in Eqs.~(\ref{relicv}) and (\ref{relica}) do not strictly hold according to our numerical calculation with \texttt{micrOmegas}. The $g_{\chi}$ values could deviate from the ones that were used in Fig.~\ref{fig:fluxes}. This will lead to overall rescalings by a factor of $1 \pm \mathcal{O}(0.1)$ for those flux curves.

\end{widetext}

\end{document}